\documentclass[reprint,amsmath,amssymb,nofootinbib,aps,pra]{revtex4-2}

\usepackage{graphicx}
\usepackage{dcolumn}
\usepackage{bm}
\usepackage[breaklinks=true,colorlinks=true,linkcolor=blue,urlcolor=blue,citecolor=blue]{hyperref}
\usepackage{multirow}
\usepackage[dvipsnames]{xcolor}
\usepackage[pscoord]{eso-pic}
\usepackage[normalem]{ulem}

\newcommand{\ie}{i.e.,~}
\newcommand{\eg}{e.g.,~}

\begin{document}
	
	
	\title{The lifetime of binary neutron star merger remnants}
	
	\author{
		Matteo Lucca,$^{1}$ and
		Laura Sagunski$^{1,2}$
		\\
		\vspace{0.15 cm}
		$^{1}$\textit{\small Institute for Theoretical Particle Physics and Cosmology, RWTH Aachen University, Aachen, 52074, Germany}\\
		$^{2}$\textit{\small Department of Physics and Astronomy, York University, Toronto, Ontario, M3J 1P3, Canada}
	}
	
	\begin{abstract}
		Although the main features of the evolution of binary neutron star systems are now well established, many details are still subject to debate, especially regarding the post-merger phase. In particular, the lifetime of the hyper-massive neutron stars formed after the merger is very hard to predict. In this work, we provide a simple analytic relation for the lifetime of the merger remnant as function of the initial mass of the neutron stars. This relation results from a joint fit of data from observational evidence and from various numerical simulations. In this way, a large range of collapse times, physical effects and equation of states is covered. Finally, we apply the relation to the gravitational wave event GW170817 to constrain the equation of state of dense matter.
	\end{abstract}
	
	\maketitle
	
	The field of gravitational-wave (GW) physics and more generally the study of the dynamics of compact objects has remained mostly theoretical since its inception. In the last four years, however, we have entered the era of GW and multi-messenger astronomy with the first GW detections of binary black hole (BBH) \cite{abbott2016a, abbott2016b, abbott2017a_BH, abbott2017f_BH, abbott2017g_BH} and binary neutron star (BNS) mergers \cite{abbott2017a_NS}.
	
	The GW emission from a BNS merger observed on August the 17th 2017, GW170817, has been particularly remarkable. In fact, in addition to the detection of GWs~\cite{abbott2017a_NS}, a worldwide collaboration of observatories followed the electromagnetic (EM) counterpart of the source \cite{abbott2017b_NS}. This innovative interplay between GW and EM detections has provided important confirmations of the theoretical framework surrounding BNS.
	
	For instance, roughly 1.7~s after the GW observation by the LIGO-Virgo collaboration, a Short Gamma-Ray Burst (SGRB) has been detected by \textit{Fermi}-GBM, GRB170817A \cite{abbott2017d_NS}. This nearly exact overlap has provided an unquestionable proof that BNS mergers are realistic sources of SGRBs. Furthermore, hours after the first burst, the UV-optical-NIR counterpart has been localized close to NGC 4993, an elliptical galaxy at a distance of approximately 40 Mpc \cite{Coulter2017}, and it was compatible with what expected by a kilonova emission \cite{abbott2017c_NS}. This observation has confirmed the belief that r-process nucleosynthesis and the subsequent products nuclear decay could take place in such environments. 
	
	However, beside the many confirmations achieved with GW170817 and GRB170817A, some questions regarding BNSs and particularly their post-merger evolution still remain open. For instance, the lifetime of one of the possible post-merger remnants, the so-called hyper-massive NS (HMNS), remains one of the characteristics with the largest uncertainties. This parameter is largely influenced by thermal effects \cite{Kaplan2013}, magnetic fields \cite{Anderson2008, Giacomazzo2011b, Kawamura:2016nmk}, viscosity \cite{Shibata2017b} and, of course, by the mass and the equation of state (EOS) of the initial NSs. In turn, the HMNS lifetime may determine the amount of mass of the torus surrounding the central star as well as its composition \cite{Fernandez2016}, and the amount of ejected matter \cite{Metzger2008, Metzger2009, Metzger2010, Wanajo2012}, which are all quantities fundamental for the modeling of the kilonova emission.
	
	Some constraints on the lifetime of HMNSs have already been determined, for instance, by means of probabilistic calculations \cite{Ravi2014}, where an upper bound at approximately ${10^4~\text{s}}$ has been found. However, in this work, we seek to find a precise relation connecting the lifetime of HMNSs to the initial parameters of the BNS system.  
	
	To investigate the free parameters of the relation, we employ two different samples. The first one has been collected from the numerical simulations of \cite{Hotokezaka2013c, Kastaun2014, Dietrich2015, Bernuzzi2015b, Kawamura:2016nmk, Rezzolla2016, Ciolfi2017, Shibata2017b, Chaurasia2018, koppel2019general}, where the HMNS collapse has happened within the simulation time. Combined, these works considered a variety of EOSs, ranging from very soft to very stiff. However, as argued in \eg \cite{Lasky2013, lu2015millisecond, gao2016constraints, piro2017fate, Radice:2017lry, Kiuchi:2019lls, Riley:2019yda, Miller:2019cac}, stiff EOSs seem to be largely disfavored with respect to EOSs of intermediate/large compactness. For this reason, in our analysis we neglect all configurations based on very stiff EOSs.
	
	As a result, these remaining simulations include 9~EOSs: ALF2 \cite{Alford2005}, APR4 \cite{Akmal1998}, BHB$\Lambda\phi$ \cite{banik2014new}, GM1 \cite{Glendenning1991Reconciliation}, GNH3 \cite{Glendenning1985}, H4 \cite{Glendenning1992}, LS220 \cite{lattimer1991generalized}, SLy \cite{Douchin2001}, and TM1 \cite{hempel2012new}. Furthermore, a large sample of physical effects is covered like, for instance, initially spinning NSs~\cite{Kastaun2014}, finite-temperature microphysical EOSs and neutrino cooling \cite{Bernuzzi2015b}, magnetized systems (\ie evolved in the framework of ideal magnetohydrodynamics) \cite{Kawamura:2016nmk, Ciolfi2017}, viscosity \cite{Shibata2017b},  eccentricity \cite{Chaurasia2018} and very short-living HMNSs \cite{koppel2019general}. Therefore, a positive overlap between their results may increase the number of possibilities the final relation can be applied to. A schematic summary of the considered configurations is displayed in Tab. \ref{tab:colltimes} (equal-mass configurations) and Tab. \ref{tab:colltimes_uneq} (unequal-mass configurations). 
	
	Note furthermore that all of these values are subject to intrinsic errors caused by the different numerical treatments. For instance, it is well know that the collapse time can be very sensitive to the chosen resolution \cite{Bauswein2010Testing, Hotokezaka2013c, Rezzolla2016}. However, we ignore these uncertainties for the rest of the discussion as they are extremely difficult to estimate.
	\begin{table}[t]
		\centering
		\caption{Summary of the properties of the simulated data for equal-mass configurations. The columns denote respectively the EOS, the mass of the initial NSs, the time of collapse of the post-merger remnants, and the reference paper.}
		\label{tab:colltimes}
		\begin{tabular}{|ccc|c|c|c|}
			\hline
			EOS & $M_{\rm NS}$ [M$_{\odot}$] & $t_{\rm coll}$ [ms] & Ref. \\
			\hline
			\small ALF2 & 1.35 & 15 & \multirow{8}{*}{\shortstack{Tab. II of \cite{Hotokezaka2013c}}} \\
			\small ALF2 & 1.40 & 5  & \\
			\small ALF2 & 1.45 & 2  & \\
			\small APR4 & 1.40 & 35 & \\
			\small H4   & 1.35 & 25 & \\
			\small H4   & 1.40 & 10 & \\
			\small H4   & 1.45 & 5  & \\
			\small SLy  & 1.35 & 10 & \\
			\hline
			\small LS220 & 1.41 & 8.6 & \multirow{2}{*}{\shortstack{Tab. II of \cite{Kastaun2014}}} \\
			\small LS220 & 1.41 & 7.7 & \\
			\hline
			\small ALF2 & 1.35 & 19 & \multirow{6}*{\shortstack{Tab. V of \cite{Dietrich2015}}} \\
			\small ALF2 & 1.35 & 19 & \\
			\small H4   & 1.35 & 22 & \\
			\small H4   & 1.35 & 25 & \\
			\small SLy  & 1.35 & 13 & \\
			\small SLy  & 1.35 & 11 & \\
			\hline
			\small LS220 & 1.35 & 48.5 & \multirow{1}*{\shortstack{Fig. 1 of \cite{Bernuzzi2015b}}} \\
			\hline
			\small H4 & 1.40 & 11.6 & \multirow{1}*{\shortstack{Tab. II of \cite{Kawamura:2016nmk}}} \\
			\hline
			\small ALF2 & 1.35  & 19.6 & \multirow{10}*{\shortstack{\cite{Rezzolla2016} \\ (From the authors)}} \\
			\small ALF2 & 1.375 & 11.7 & \\
			\small ALF2 & 1.40  & 6.8  & \\
			\small GNH3 & 1.35  & 24.2 & \\
			\small GNH3 & 1.375 & 11.8 & \\
			\small GNH3 & 1.40  & 10.2 & \\
			\small H4   & 1.375 & 21.4 & \\
			\small H4   & 1.40  & 14.7 & \\
			\small SLy  & 1.375 & 3.9  & \\
			\small SLy  & 1.40  & 2.8  & \\
			\hline
			\small H4 & 1.35 & 22 & \multirow{1}*{\shortstack{Fig. 24 of \cite{Ciolfi2017}}} \\
			\hline
			\small H4 & 1.30 & 14 & \multirow{1}*{\shortstack{Fig. 4 of \cite{Shibata2017b}}} \\
			\hline
			\small SLy & 1.36 & 6.4  & \multirow{2}*{\shortstack{Tab. III of \cite{Chaurasia2018}}} \\
			\small SLy & 1.36 & 29.6 & \\
			\hline
			\small BHB$\Lambda\phi$ & 1.53  & 0.3 & \multirow{7}*{\shortstack{Fig. 2 of \cite{koppel2019general}}} \\
			\small BHB$\Lambda\phi$ & 1.55  & 0.3 & \\
			\small BHB$\Lambda\phi$ & 1.57  & 0.2 & \\
			\small TM1 & 1.65  & 0.5 & \\
			\small TM1 & 1.65  & 0.4 & \\
			\small TM1 & 1.67  & 0.3 & \\
			\small TM1 & 1.68  & 0.3 & \\
			\hline
		\end{tabular}
	\end{table}
	
	\begin{table}[t]
		\centering
		\caption{Summary of the properties of the simulated data for unequal-mass configurations. The columns denote the same quantities as in Tab. \ref{tab:colltimes} with the addition of the mass-ratio $q$. $M_{\rm NS}$ refers to the more massive NSs of the binary.}
		\label{tab:colltimes_uneq}
		\begin{tabular}{|cccc|c|c|c|}
			\hline
			EOS & $M_{\rm NS}$ [M$_{\odot}$] & $t_{\rm coll}$ [ms] & $q$ & Ref. \\
			\hline
			\small ALF2 & 1.50 & 45 & 0.80 & \multirow{9}{*}{\shortstack{Tab. II of \cite{Hotokezaka2013c}}} \\
			\small ALF2 & 1.45 & 40 & 0.86 & \\
			\small ALF2 & 1.40 & 10 & 0.93 & \\
			\small APR4 & 1.50 & 30 & 0.87 & \\
			\small H4   & 1.60 &  5 & 0.81 & \\
			\small H4   & 1.50 & 20 & 0.87 & \\
			\small SLy  & 1.50 & 10 & 0.80 & \\
			\small SLy  & 1.45 & 15 & 0.86 & \\
			\small SLy  & 1.40 & 15 & 0.93 & \\
			\hline
			\small SLy & 1.45 & 15 & 0.86 & \multirow{4}*{\shortstack{Tab. V of \cite{Dietrich2015}}} \\
			\small SLy & 1.45 & 14 & 0.86 & \\
			\small SLy & 1.45 & 16 & 0.86 & \\
			\small SLy & 1.45 & 11 & 0.86 & \\
			\hline
			\small LS220 & 1.44 & 5 & 0.97 & \multirow{1}*{\shortstack{Fig. 1 of \cite{Bernuzzi2015b}}} \\
			\hline
			\small H4 & 1.54 & 24.7 & 0.8 & \multirow{1}*{\shortstack{Tab. II of \cite{Kawamura:2016nmk}}} \\
			\hline
		\end{tabular}
	\end{table}
	
	All of these studies only include configurations where the collapse occurs approximately in the first 50 ms after the merger. It is however likely (see \eg \cite{Ravi2014}) that the lifetime of HMNSs may reach values up to $10^4$ $\rm s$. Therefore, in order to find a behavior covering the whole range of collapse times, we extend our results by considering the calculations of \cite{Lasky2013} and \cite{lu2015millisecond}. Both works are based on the observational results of the \textit{Swift} satellite~\cite{Gehrels_2004} and the value for the collapse time is obtained by fitting the observed light curves using a simplified model where dipole electromagnetic radiation is the main mechanism of angular momentum loss. The considered data points are summarized in Tab. \ref{tab:colltimes 2}. Note that this data set is fundamentally different from the one discussed above, not being the result of numerical simulations and, as clear from the references, all values for the mass are subject to uncertainties up to ${0.03~\text{M}_{\odot}}$.
	\begin{table}[t]
		\centering
		\caption{Summary of the properties of the observational data. The columns denote the same quantities as in Tab. \ref{tab:colltimes} (including the assumption of $q=1$).}
		\label{tab:colltimes 2}
		\begin{tabular}{|ccc|c|c|c|}
			\hline
			EOS & $M_{\rm NS}$ [M$_{\odot}$] & $t_{\rm coll}$ [s] & Ref. \\
			\hline
			\small APR4 & 1.10 & 92 & \multirow{12}*{\shortstack{Tab. I and Fig. 1 \\ of \cite{Lasky2013}}} \\
			\small APR4 & 1.10 & 343 & \\
			\small APR4 & 1.09 & 291 & \\
			\small APR4 & 1.13 & 139 & \\
			\small GM1 & 1.20 & 92 & \\
			\small GM1 & 1.20 & 343 & \\
			\small GM1 & 1.18 & 291 & \\
			\small GM1 & 1.13 & 139 & \\
			\small SLy & 1.03 & 92 & \\
			\small SLy & 1.03 & 343 & \\
			\small SLy & 1.03 & 291 & \\
			\small SLy & 1.06 & 139 & \\
			\hline
			\small APR4 & 1.12 & 118 & \multirow{24}*{\shortstack{Fig. 11 of \cite{lu2015millisecond}}} \\
			\small APR4 & 1.12 & 134 & \\
			\small APR4 & 1.12 & 61 & \\
			\small APR4 & 1.12 & 52 & \\
			\small APR4 & 1.12 & 53 & \\
			\small APR4 & 1.13 & 146& \\
			\small APR4 & 1.12 & 137 & \\
			\small APR4 & 1.12 & 114 & \\
			\small GM1 & 1.20 & 118 & \\
			\small GM1 & 1.20 & 134 & \\
			\small GM1 & 1.20 & 61 & \\
			\small GM1 & 1.20 & 52 & \\	
			\small GM1 & 1.20 & 53 & \\
			\small GM1 & 1.21 & 146 & \\
			\small GM1 & 1.20 & 137 & \\
			\small GM1 & 1.20 & 114 & \\
			\small SLy & 1.04 & 118 & \\
			\small SLy & 1.04 & 134 & \\
			\small SLy & 1.04 & 61 & \\
			\small SLy & 1.04 & 52 & \\
			\small SLy & 1.04 & 53 & \\
			\small SLy & 1.05 & 146 & \\
			\small SLy & 1.04 & 137 & \\
			\small SLy & 1.04 & 114 & \\
			\hline	
		\end{tabular}
	\end{table}
	
	Our goal is now to find a relation connecting the lifetime of the HMNS, $t_{\rm coll}$ (here we employ the common convention of setting $t_{\rm merger}=0$~s), and the initial parameters of the BNS system. 
	
	Since the collapse times can range over several orders of magnitude, it is natural to assume a power-law dependence on the masses of the NSs and the corresponding radii, which are the most characteristic quantities describing and influencing the evolution of the initial NSs. In the following, we express the product of the NSs masses, $M_{\rm NS,1}M_{\rm NS,2}$, as $qM_{\rm NS,1}^{2}$, where $q=M_{\rm NS,2}/M_{\rm NS,1}$ is the mass ratio. For simplicity, we also assume that $M_{\rm NS,1}\equiv M_{\rm NS}$ is the more massive configuration. As a result, we obtain a proportionality of the simple form
	\begin{align}\label{eq:t_coll 1}
	\frac{t_{\rm coll}}{t_0}\propto\left(\frac{M_{\rm NS}\sqrt{q}}{M_{\rm TOV}}\right)^{e_1}\left(\frac{R_{\rm NS,1}R_{\rm NS,2}}{R_{\rm TOV}^{2}}\right)^{e_2}\,,
	\end{align}
	where, in order to only have dimensionless quantities involved in the relation, we normalize the collapse time with a free parameter $t_0$, the NS mass with the TOV mass\footnote{Throughout the paper, the subscript $_{\rm TOV}$ indicates a quantity referring to the maximum mass (non-rotating) neutron star model for a given EOS, i.e. the Tolman-Oppenheimer-Volkoff (TOV) limit.}, $M_{\rm TOV}$, predicted by the given EOS, and the radius with $R_{\rm TOV}$. The exponents $e_1$ and $e_2$ are two free parameters yet to be determined ($e_1$ also includes a factor 2 due to the product of the masses).
	
	Physically, we would expect the collapse time to be anti-proportional to the initial NS mass, as more massive systems are expected to collapse faster, thus lowering the value of $t_{\rm coll}$. This can be used as a consistency check, once the free parameters $e_1$ and $e_2$ have been determined. Furthermore, this expectation also allows us to interpret the value of $t_0$ as a definition of the minimum lifetime of a HMNS. In fact, Eq. \eqref{eq:t_coll 1} reveals that, considering mass configuration with the shortest lifetime, \ie assuming the maximum mass configuration for a given EOS, the ratio $t_{\rm coll}/t_0$ is equal to one. Note as a further remark that the value of $t_0$ is universal, \ie independent of the chosen~EOS. 
	
	Taking the logarithm of both sides of Eq. \eqref{eq:t_coll 1}, directly predicts a linear behavior of type
	\begin{align}\label{eq:t_coll_fit 1}
	\log(t_{\rm coll})= e_0+ e_1 \log & \left[ \frac{M_{\rm NS}\sqrt{q}}{M_{\rm TOV}} \left(\frac{R_{\rm NS,1}R_{\rm NS,2}}{R_{\rm TOV}^{2}}\right)^{e_2/e_1}\right],
	\end{align}
	where $e_0=\log(t_0)$, $e_1$  and $e_2$ represent the different degrees of freedom (DOF) involved in the relation.
	
	We validate the relation in Eq. \eqref{eq:t_coll_fit 1} by confronting it against the data samples given in Tabs. \ref{tab:colltimes}, \ref{tab:colltimes_uneq} and \ref{tab:colltimes 2}. To determine the best fit values of the free parameters of the model, \ie $\{e_0,e_1,e_2\}$, we perform a Markov Chain Monte Carlo (MCMC) scan over the parameter space. Interestingly, our MCMC results indicate a strong preference for for the ratio $e_2/e_1$ to be very close to zero, $e_2/e_1=-0.09$, pointing towards a radius-independent relation.
	
	To further investigate this aspect, we run a second set of MCMCs removing the radius dependence in Eq. \eqref{eq:t_coll_fit 1}. We thus obtain
	\begin{align}\label{eq:t_coll_fit 2}
	\log(t_{\rm coll})= e_0+e_1 \log\left(\frac{M_{\rm NS}\sqrt{q}}{M_{\rm TOV}}\right)\,,
	\end{align}
	and only treat $\{e_0,e_1\}\,$ as free parameters.
	
	Note that a similar proportionality has already been proposed in \cite{koppel2019general}. There, however, the collapse time has been normalized with the free-fall time for the maximum-mass model, which introduces an additional dependence on $R_{\rm TOV}$. In our relation, this factor is not explicitly present, but its impact might be absorbed in $t_0$, which plays an analogous role in Eq. \eqref{eq:t_coll 1} and hence in Eq. \eqref{eq:t_coll_fit 2}.
	
	The best fit values for the free parameters $\{e_0,e_1\}$ in the case of Eq. \eqref{eq:t_coll_fit 2} are given by $e_0=-5.51 \pm 0.36$ and $e_1=-39.0 \pm 1.6$, where the errors represent the 1$\sigma$ confidence levels. In comparison to before, the best fit values for $e_0$ and $e_1$ change only marginally (by less than a factor of 1.7 and $0.48 \%$, respectively). Additionally, the absolute value of the $\chi^2$ only increases by $15 \%$, yielding a value of ${\chi^2=12.9}$. Furthermore, the errors of the free parameters, in particular of $e_0$, are smaller when the value of $e_2$ is fixed. We can thus safely assume that Eq. \eqref{eq:t_coll_fit 1} is indeed independent of the radius and can be reduced to the simple form of Eq. \eqref{eq:t_coll_fit 2}.
	\begin{figure}[t]
		\centering
		\includegraphics[width=85mm, height=65 mm]{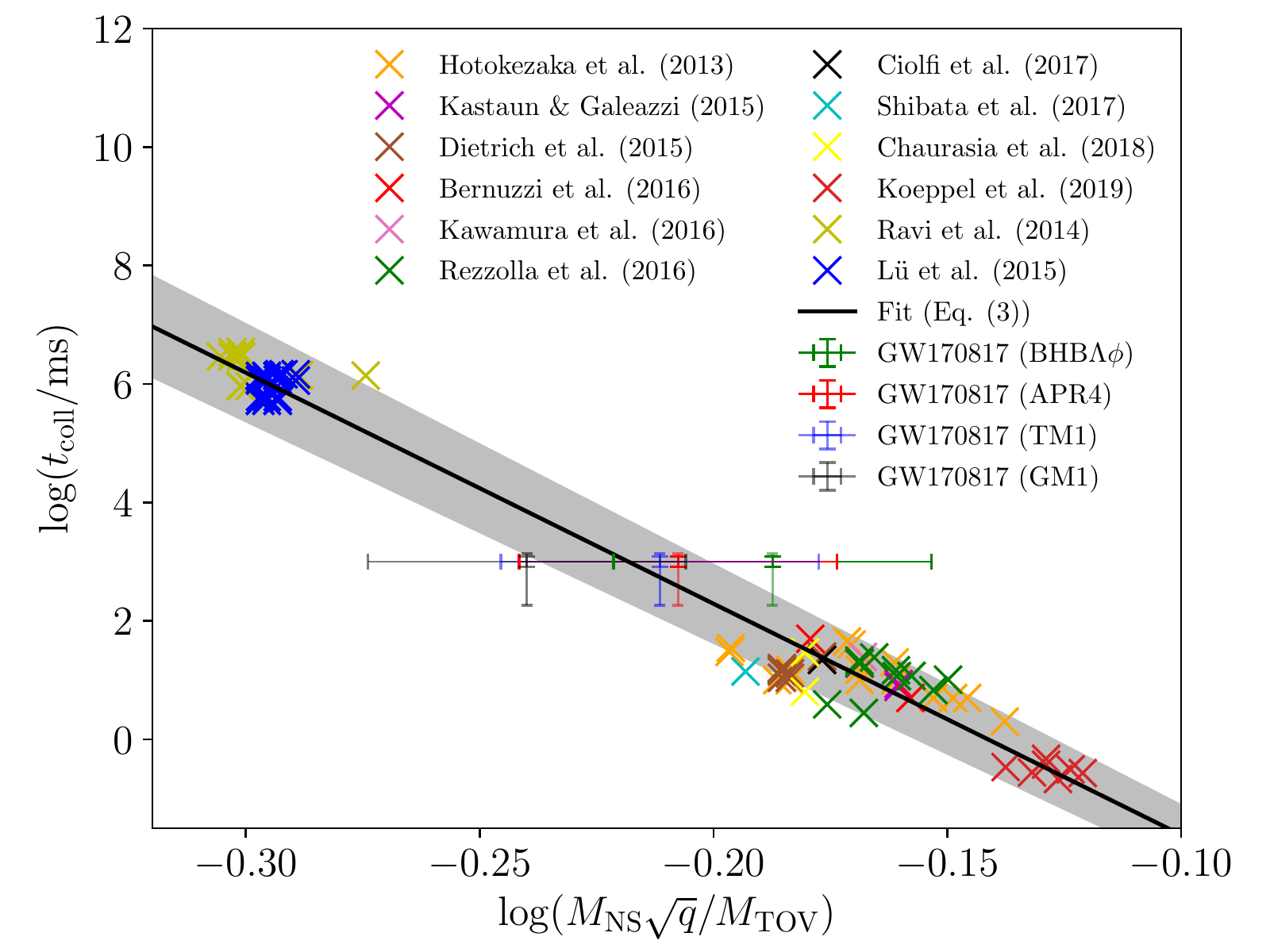}
		\caption{Collapse time $t_{\rm coll}$ as function of $(M_{\rm NS}/M_{\rm TOV})$. The crosses represent the predictions of the references listed in Tabs. \ref{tab:colltimes}, \ref{tab:colltimes_uneq} and \ref{tab:colltimes 2}. The black line corresponds to Eq.~\eqref{eq:t_coll_fit 2} using the best fit values for $e_0$ and $e_1$, while the gray band corresponds to the 1$\sigma$ contour. To illustrate the case of GW170817, we display the bounds on the mass and the lifetime as horizontal and vertical lines, respectively, for 4 representative EOSs (BHB$\Lambda\phi$, APR4, TM1 and GM1). These 4 points are not included in the fitting procedure.}
		\label{fig:t_coll}
	\end{figure}
	\begin{figure*}[t!]
		\centering
		\includegraphics[width=\textwidth]{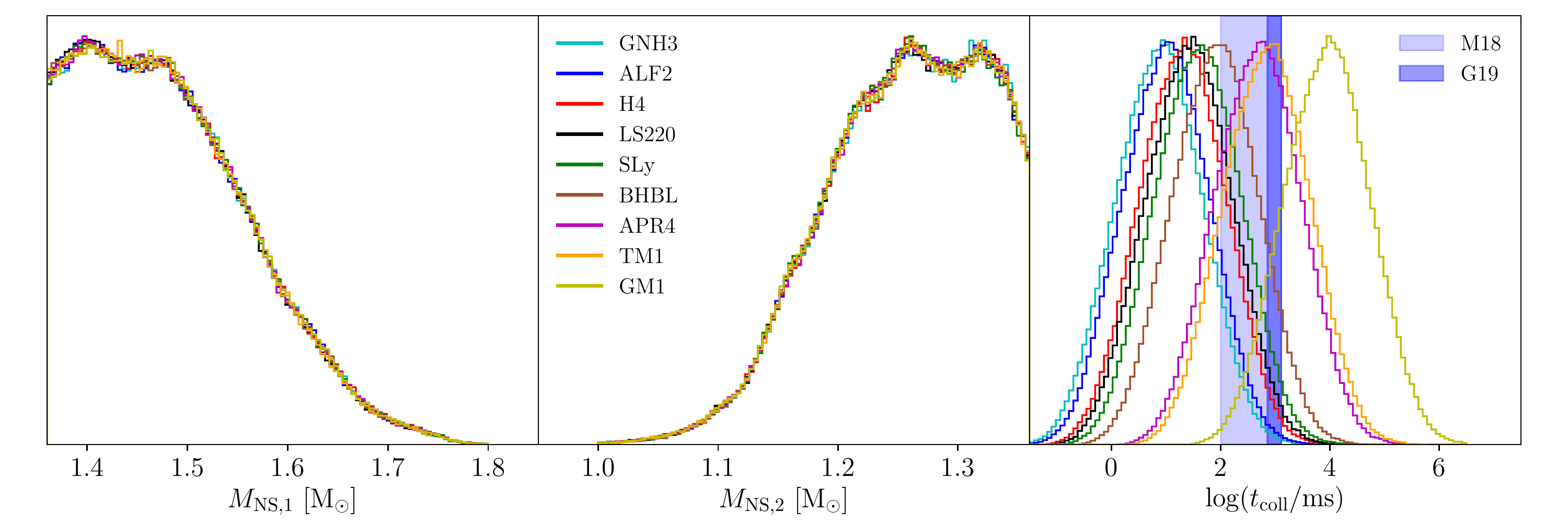}
		\caption{Probability distributions (from left to right) of the initial NS masses of GW170817 as given in \cite{abbott2017a_NS} and of the corresponding HMNS lifetime for all EOSs considered in this work, as predicted by Eq.~\eqref{eq:t_coll_fit 2}. The horizontal blue bands in the right panel correspond to the constraint set by the kilonova emission (here labeled M18 \cite{Metzger_2018} and G19 \cite{gill2019did} for shortness).}
		\label{fig:GW170817_constraint}
	\end{figure*}

	As a remark, note that the best fit values might be partly biased by the limited sample of EOSs and, more probably, by the non-homogeneous distribution of the points among the many EOSs. However, quantitatively evaluating the presence and the strength of this bias will only be possible once additional data becomes available.
	
	Once the value of $e_0$ has been determined, it is thus possible to compute the minimum lifetime of a HMNS as $t_0\approx 3$ ns. This value can be interpreted as a formal definition of the benchmark below which the binary system cannot form a HMNS in its post-merger evolution and promptly collapses to a BH.
	
	In Fig. \ref{fig:t_coll} we show the data points for the collapse time $t_{\rm coll}$ as function of the normalized mass $M_{\rm NS}/M_{\rm TOV}$ for the many configurations listed in Tabs. \ref{tab:colltimes}, \ref{tab:colltimes_uneq} and \ref{tab:colltimes 2}. There, also the fitting relation expressed in Eq. \eqref{eq:t_coll_fit 2} is displayed as a solid black line together with its 1$\sigma$ bounds as gray band. As expected, for very large NS masses, the HMNS lifetime decreases.
	
	Observing the relation shown in Fig.~\ref{fig:t_coll}, however, a limitation of our datasets becomes evident. In fact, since the sources of our data points are either time-limited simulations or broad SGRB plateaus, the predicted HMNS lifetimes can only be very short or very long, respectively. This creates a gap in the time range between 50 ms and 50 s where the validity of our linear relation might be uncertain.
	
	In order to fill this gap, we additionally consider the observational data of GW170817. Indeed, detailed analyses \linebreak of the kilonova emission following GW170817 and its afterglow have allowed to establish an approximate lifetime interval for the HMNS that could explain the observed light curves and spectra. According to \cite{Metzger_2018}, it spans approximately from 100 ms and to 1 s, while the more advanced calculations performed in \cite{gill2019did} narrowed the range down to 0.72--1.29 s. The latter constraints were found by requiring that the merger remnant sustains both the observed blue-ejecta mass and the relativistic jet necessary to power up the emission. These estimated lifetimes lay exactly in the center of the gap left by our samples and can thus give a strong indication for the validity of our relation also at these scales.
	
	Furthermore, in the analysis conducted by the LIGO-Virgo collaboration \cite{abbott2017a_NS, abbott2017b_NS, abbott2017c_NS, abbott2017d_NS}, two possible configurations for the initial state of GW170817 where presented. Depending on the assumption made about the spin of the initial NSs, each scenario predicts a different combination for the initial NS masses. In the physically more plausible case modeled by a low-spin prior (spin restricted to $\leq 0.05$) \citep{abbott2017d_NS}, the masses $M_{\text{NS},1}$ and $M_{\text{NS},2}$ of the single NSs range between (1.36, 1.60) M$_{\odot}$ and (1.17, 1.36) M$_{\odot}$, respectively, while the total mass is $2.74^{+0.04}_{-0.01}$ M$_{\odot}$ \cite{abbott2017a_NS}.
	
	Making an assumption on the EOS, it is thus possible to display the prediction for GW170817 in Fig.~\ref{fig:t_coll}. There, the horizontal and vertical thick lines correspond to the bounds on the mass and the lifetime given for GW170817 in \cite{abbott2017a_NS} and \cite{gill2019did}, respectively, while the thin vertical lines correspond to the more conservative bounds derived in~\cite{Metzger_2018}. As representative examples, we select the EOSs BHB$\Lambda\phi$, APR4, TM1 and GM1. This choice is based on several previous analyses \cite{Lasky2013, lu2015millisecond, gao2016constraints, piro2017fate, Radice:2017lry, Kiuchi:2019lls, Riley:2019yda, Miller:2019cac} and on the final results of this work. As one can see in Fig. \ref{fig:t_coll}, the relation found in Eq. \eqref{eq:t_coll_fit 2} seems to be nicely compatible with these choices, although the current bounds on the mass are not very stringent. We can thus safely extend our relation also to mid-long HMNS lifetimes. Note, however, that these 4 points are not included in the fitting procedure.
	
	Finally, Eq. \eqref{eq:t_coll_fit 2} does not only represent a single compact relation connecting the initial state of a BNS system with the collapse time of its post-merger remnant, but it also possesses a remarkable constraining power.
	
	To mention one of the possible applications of Eq. \eqref{eq:t_coll_fit 2} to constrain the NS EOS, we focus our attention again on GW170817. In fact, in this case it is possible to employ the initial NS mass distributions given in \cite{abbott2017a_NS} for GW170817 to compute the corresponding probability distribution for $t_{\rm coll}$ using Eq. \eqref{eq:t_coll_fit 2}. By comparing these predictions to the results of \cite{Metzger_2018, gill2019did} for every EOS, one can then draw conclusions on the probability of a given EOS to reproduce the data.
	
	The results of this procedure are shown in Fig. \ref{fig:GW170817_constraint}. In the first two panels from the left, we plot the posterior distribution of the initial NS masses of GW170817 as given in e.g., Fig. 4 of \cite{abbott2017a_NS}. In order to ensure that the combination of the two masses adds up to the total of $2.74$~M$_{\odot}$, we impose a prior on the mass-ratio $q$ following the results presented e.g., in Fig. 7 of \cite{Abbott2018_NS}. In the right panel of the figure, the different EOS-dependent distributions for $t_{\rm coll}$ are displayed. In the same panel, the horizontal blue bands represent the constraint set by the kilonova emission (dark blue for \cite{gill2019did} and light blue for \cite{Metzger_2018}).
	
	Fig. \ref{fig:GW170817_constraint} shows in particular that stiff EOSs like GNH3 are far from the acceptance region, while softer EOSs like BHB$\Lambda\phi$, APR4 and TM1 seem to be more realistic.~This nicely confirms what already found in the literature~\cite{Lasky2013, lu2015millisecond, gao2016constraints, piro2017fate, Radice:2017lry, Kiuchi:2019lls, Riley:2019yda, Miller:2019cac}. Furthermore, future, soon-to-be-expected detections of gravitational signals from BNSs (should they also have an observable EM counterpart) will allow us to both confirm the validity of the relation found in this work and constrain the EOS of NSs even further.
	
	In addition to its constraining power, our relation has various other interesting applications. For instance, it can be used in a more complex analysis of the BNS evolution to extract information on the gravitational emission, its electromagnetic counterpart and the combination of the two. As an example, it would be very interesting to explore the connection between the time evolution of the angular momentum of the HMNS and its lifetime in future work. 
	
	On top of that, the relation can be used as a very simple and efficient tool for the setup of time-consuming numerical simulations to forecast the lifetime of the post-merger remnant. This will significantly facilitate the choice of masses and EOSs.
	\\
	\\
	\textit{Acknowledgments}: The authors thank C. Fromm, F. Guercilena, M. Johnson, T. Konstandin, J. Lesgourgues, P. Mertsch, L. Rezzolla, N. Sch\"{o}neberg and S. Tulin for helpful discussions and numerous useful comments. The authors especially want to thank the referee for the many constructive suggestions. The authors also thank K. Takami for providing access to his simulation data. LS is supported by the Natural Sciences and Engineering Research Council of Canada.
	
	\bibliography{bibliography}
	
\end{document}